# Statistical analysis of the price and subjective quality ratings on Australian wines

*Peter M. Visscher*

Queensland Brain Institute, The University of Queensland, Building 79, St Lucia 4072

**Abstract**
Consumers have a wide choice regarding the purchase of wine. Wines vary across varieties, regions, years of vintage, alcohol content, price and expert ratings. In this study we explored the relationship between those variables and whether a combination of explanatory variables can predict outcome variables using data on selected 2710 Australian wines rates by the latest book of James Halliday[1]. We analysed two dependent (outcome) variables, namely the rating and the price, and their (non-linear) relationship between explanatory variables and each other. We observe that the rating scale is not linear, as consumers may believe, and that a narrow range of rating scores is used for the ascertained wines. Across all wines, approximately 20% of variation in log(Price) can be explained by the explanatory factors state, variety, vintage and alcohol percentage. If the residuals from the model for rating is also taken as an explanatory variable for price, then a total of 53% of variation in log(Price) is explained by the model. For those wines ascertained to be in Halliday[1], there is more variability between white wines than between red wines that is explained by explanatory variables such as State and variety. In comparison with hedonic pricing analyses in previous studies on French and Australian wines, a smaller proportion of variation in price and rating was explained in Australian wines that were in the 2014 edition of Halliday.

**Introduction**

Consumers have a wide choice regarding the purchase of wine, with the availability of individual wines across varieties, districts, years of vintage, alcohol content, price and expert ratings. In this study we explored the relationship between those variables and whether a combination of explanatory variables can predict outcome variables using data on Australian wines. We assumed that there are two dependent (outcome) variables, namely the rating and the price. Our aims were firstly, to explain variation in these variables from explanatory variables such as variety, geographical region, year of vintage and alcohol percentage and secondly to determent the function that best explains the relationship between rating and price. That is, if rating and price are correlated, what scale of measurement best captures this relationship and what is the strength of association?

**Data**

Data consisted of a selection of wines from the 2014 edition of the Australian Wine Companion [1]. For each selected entry, the state of the winery, variety, year of vintage, alcohol percentage, rating and price were recorded only if there were no missing data. On some wines there was no alcohol percentage recorded and these were not considered for analyses. Some wines were scored by raters other than Halliday and these were not taken into account, so that all scores were from a single rater. Only red and white wines of the major varieties were considered listed here with their acronyms: Red wines, Shiraz (SH), Cabernet Sauvignon (CS), Pinot Noir (PN), Merlot (ME) and Red Blend of any varieties (RB). White wines, Sauvignon Blanc (SB), Chardonnay (CH), Riesling (RI), Pinot Grigio or Pinot Gris (PG) and White Blend of any varieties (WB). Data were manually entered. Sweet wines, sparkling wines, 100% Semillon, rose and non-vintage wines were not considered.

According to Halliday[1], the price (in Australian dollars) is the amount is specified by the winery and may not reflect the price consumers pay. Subjective ratings are on a scale from 75-100, with wines in the range 94-100 described as 'outstanding', 90-93 'Highly recommended', 87-89 'Recommended', 84-86 'Acceptable', 80-83 'Over to you' and 75-79 'Not recommended'.

**Data analysis**

All statistical analyses were performed using the free statistical package "R"[2]. Summary statistics were produced across states, varieties and years of vintage. Price shows a strong skewed distribution (Supplementary Figure 1). A Box-Cox transformation indicated that $1/\sqrt{}$(Price) is approximately the best transformation in a regression of Price on Alcohol percentage with regard to the distribution of residuals. However, a more intuitive log(Price) has a correlation of -0.98 with $1/\sqrt{}$(Price) and for subsequent analyses we have used log2(Price) for ease of interpretation of results, consistent with analysis of price or income in economics (including previous analyses of wine prices) where a logarithmic transformation is common. In our notation, log(y) is used when the base of the logarithm is irrelevant (since logarithms of different bases are proportional to each other), and the base is given when the scale matters. The distribution of the log(log(Price)) appears slightly more symmetrical than that of log(Price) but analysis results using either of these transformations were very similar and we used log2(Price) for all analyses, unless stated explicitly.

**Results**

**Summary statistics**

A total of 2710 wines fulfilled the selection criteria. Figures 1-3 give a breakdown of the wines by variety and states. Shiraz (29%) and Chardonnay (15%) dominate the varieties, and 81% of all entries are from South Australia, Victoria and Western Australia. Figure 4 shows the distribution of wines over year of vintage. The majority of wines are from vintages 2010, 2011 and 2012.

Figure 5 shows the distribution of alcohol percentage across all wines. However, there is a difference in the distribution across varieties and Table 1 summarises the alcohol percentage, rating and price across varieties. The average alcohol percentage of the red wines is significantly larger than that of the white wines, with Cabernet Sauvignon, Red Blends and Shiraz varieties averaging more than 14%. Mean ratings are in the range of 91 to 93, with the largest values of 93.5 and 93.3 for Riesling and Shiraz, respectively.

Figure 6 shows the distribution of log2(Price) by Variety. The distribution of log(Price) appears reasonably symmetrical.

Rating is the outcome of a strong ascertainment procedure – for example the lowest value is 87 yet the scale is defined in the range of 75-100. The histogram of Rating by Variety shows a non-standard distribution, with a spike at Ratings of 93 (Figure 7). For analysis we used log2(Ratings) which has a more symmetrical distribution (Figure 8). It also aids with interpretation of the regression analyses.

**Model fitting**

*Explaining variation in Rating.*

A linear model with log2(Rating) as dependent variable and Year, State, Variety and Alcohol was fitted. The model explained 12% of variation in log2(Rating). At a significance level of $p < 0.001$, several States and Varieties were statistically significant. Merlot and Pinot Grigio had the lowest estimates of log(Rating) whereas Riesling had the highest estimate. Of the States, WA and SA had the highest estimated effect size. Alcohol was significant ($p = 0.0001$), with an estimated effect size of 0.004 log2(Rating) per 1% increase in alcohol. However, the effect size is small, with an increase in Rating of $2^{0.004} \sim 1.003$ per 1% increase in alcohol content, an increase in of approximately 0.3 points on the observed rating scale per 1% alcohol increase.

Splitting up the analysis by red and white wines shows that more variation in log(Rating) is explained for white wines (18%) than for red wines (7%), see Table 2. When the four varieties with the largest sample size (SH, RB, CH and RI) are analysed individually, the model explains 4% (SH, RB) to 13% of variation in log(Rating) (Table 2). For individual varieties, State, Alcohol and Year were fitted as explanatory variables.

*Explaining variation in Price*

The same model as for log2(Rating) was fitted for log2(Price). The model explained 20% of variation in log(Price). [When modelling log(log(Price)) instead of log(Price), 20.8% of variation was explained by the linear model]. Of the significant Varieties, Merlot and Sauvignon Blanc had the lowest estimates and Shiraz the largest. Alcohol was highly significant ($p = 4.10^{-8}$), with an estimated effect of 0.125 in log2(Price) per percent alcohol. Hence, after adjustment for all other explanatory variables, an increase of 1% alcohol is associated with a predicted proportional increase of $2^{0.125} = 1.09$ in the price, so approximately a 10% increase in the price per 1% increase in alcohol percentage.

As for Rating, more variation among white wines is explained by the model than for red wines, 27% and 11% in log(Price) respectively (Table 2). For individual varieties, the proportion of variation in log(Price) explained by the model ranged from 4% (RI) to 11% (SH) (Table 2).

*Relationship between Price and Rating*

The Pearson correlation between log2(Price) and log2(Rating) across all wines was 0.64, and the correlations between log2(Rating) and Alcohol 0.11 and between log2(Price) and Alcohol of 0.30. However, these correlations could be confounded with variety and year. The residuals from the previously fitted linear models of log2(Price) and log2(Rating) are plotted in Figure 6. The correlation between these residuals is 0.65. Hence, approximately 42% ($0.65^2$) of variation in log(Price) is explained by log(Rating) and vice versa. A linear regression analysis of the residuals of log2(Price) on the residuals of log2(Rating) has a slope of 12.8, so approximately a 13% increase in price per 1% increase in rating.

The correlation between residuals for log(Rating) and log(Price) estimated from separate analyses of red and white varieties was 0.64 and 0.65 separately.

**Discussion**

We have analysed the entries of one specific rater of all Australian wines that were listed in the 2014 edition of a wine book and satisfied the inclusion criteria in this study. The presented summary statistics indicate known differences between states, varieties and price. What general conclusion can we draw from the statistical analyses?

Firstly, the rating scale is not linear, as some consumers may believe. Clearly an increase from a rating of 93 to 94 is not the same as an increase from 94 to 95. It is surprising what a narrow range of rating scores are used, although this may be a reflection of ascertainment if wines with lower scores did not appear in the printed version. Secondly, across all wines, approximately 20% of variation in log(Price) can be explained by the explanatory factors state, variety, vintage and alcohol percentage. If the residuals from the model for rating is also taken as an explanatory variable for price, then a total of 53% of variation in log(Price) is explained by the model. Similarly, if the residuals from the model for price is used as an explanatory variable for rating, then 48% of variation in log(Rating) is explained by the model. However, for these comparisons it is questionable whether rating or price should be used as an explanatory variable. Thirdly, for those wines ascertained to be in Halliday[1], there is more variability between white wines than between red wines that is explained by explanatory variables such as State and variety (Table 3).

A correlation of ~0.65 between the residuals of log(Price) and log(Rating) appears large but gives the opportunity to identify wines that have a favourable Rating/Price ratio, after adjusting for known explanatory variables such as variety, State and alcohol percentage.

Quantitative analyses of wine prices and ratings have been performed before, on Bordeaux wines [3-5], Australian wines [6-8] and other wines (e.g.[9]) . Most of these studies are what economists call 'hedonic pricing' analyses. In hedonic pricing analysis methods, it is assumed that the price of good or service depends both on internal and external factors, and the aim is to identify and quantify factors that affect the price. In Table 3 we give an summary of model fitting results from previous analyses.

Oczkowski[7] used data on 1604 Australian premium wines from the 1991 and 1992 edition of a wine book by Shield and Meyer. A log-linear was fitted, in which log(Price) was modelled as a linear function of quality rating, cellaring potential, grape variety, grape region, grape vintage, producer size, year of marketing and all interactions (at total of 105 effects were estimated). The adjusted $R^2$ from the model was 58%, and this includes the effect of quality rating. For the effect of grape variety of table wines, 'Light Reds' and 'Rhine Riesling' has the lowest score whereas Pinot Noir and Cabernet Sauvignon/Merlot the highest. The results of the present study indicate that much has changed since 1992, since Riesling now has a positive effect on price. However, this comparison may be biased because 'Rhine Riesling' in the early 1990s could indicate a range of grape varieties, not necessarily Riesling.

Schamel and Anderson[10] analysed data from wines of Australia and New Zealand over the vintages 1992-2000. The data were from two sources, the James Halliday books for each vintage from 1992 to 2000 (n = 6866 observations for Australian wines), and the Winestate magazine ratings for 11,251 Australian wines from the 1992 to 1999 vintages. The authors use the log(Price) scale, consistent with most other studies, including the present one. Their model included 27 Australian regions, variety and rating (both a 'vintage' and a 'winery' rating. The vintage rating is similar in scale and purpose as the rating used in the current study).  For Australian wines from the Halliday data, they report a 2.3 to 4.1 per cent increase for a one-point increase in rating quality, amounting to an estimated effect of 41 to 110 Australian cents per 1-point quality rating. The model in total explained 29 to 45% of variation in log(Price). However, this includes the effect of rating and it is not clear that this variable should be used as an explanatory factor, for example it cannot be used to make predictions because ratings just like price are outcome variables. Nevertheless the authors conclude that "ratings appear to have a significant positive impact on the prices that consumers are willing to pay".

Wood and Anderson[8] used auction price data from four icon wines in Australia from vintages of 1971 to 1993. The age of the wine, several weather variables and wine technique factors were tested for association with log(Price). Excluding age, the adjusted $R^2$ of the regression models ranged from 45 to 67% for the four wines, increasing to 59-78% when age was included.

For French wines, Landon and Smith[5] investigated the role of 'collective reputation' (past performance) in explaining variation in the price of 559 observations on 196 different red wines from Bordeaux vintages of 1987 to 1991. For the functional form they found that the reciprocal square root model was best (as found in the present study) and used that in subsequent model fitting. Their full model included 38 variables, including region, quality classification scores, grape variety and the size of the producer. The model also included the effect of an indicator of collective reputation, defined as a score given by the well-known wine rater Robert Parker. The full model explained 75% of variation in the reciprocal square root of the price. The authors conclude that (collective) reputation is an important determinant of what consumers pay for wine and that many hedonic price studies may overstate the effect of current quality on price. Combris et al.[4] analysed data from an experimental study in which juries evaluated and scored samples of a number of a Bordeaux wines. Since this was an experiment, the authors could separate the effects of 'label' characteristics (such as vintage year, name of producer) and sensory characteristics (taste, texture, odour). Data were on 519 wine bottles from three Bordeaux vintages (1989-1991). A model that included sensory and objective variables explained 63% of variation in log(Price), but most selected variables were label characteristics. The authors also analysed the jury grade as an outcome variable and explained 60% of variation of log(Rating). The effect of alcohol percentage was negative, so that the jury grade decreased for increasing alcohol percentage. In contrast to the model for price, the selected variables in the model for jury rating were mostly sensory variables. The correlation between price level and jury grade was 0.44, but when the authors tested for a correlation between the residuals of the two regression models for price and rating, their results were consistent with independence. They concluded that after conditioning on the explanatory variables the price and jury grade were independent. This is clearly in contrast to the current study where the residuals were highly correlated (r = 0.65). A hedonic approach to predict the quality and price of mature Bordeaux wines from selected chateaux from data of vintages 1952 to 1980 was used by Ashenfelter[3]. Variation in log(Price) was regressed against age of vintage, the amount of rain in August and the amount of rain in the months preceding the vintage. Age alone explained 21% of variation in log(Price) whereas adding the weather variables increased that to 83%. A predictor from the model worked well for out-of-sample data. For Bordeaux wines, there is evidence that expert opinion that is independent to the fundamental determinants of the quality of wine has an influence on the price in the short run [3].

The amount of variation in price of the current study seems low when compared to other studies (Table 3). This is likely because other studies either included expert rating as an explanatory variable or they used label characteristics such as the winery into account. The choice of which samples to include in the analysis may also be important. Some studies have a narrow range of wines (e.g. those from a few producers) and capture variation between years due to the weather, other studies have ascertained a wider range of wines in terms of price and (perceived) quality, thereby increasing variation in the sample. In terms of model fitting, the main difference between studies is whether to include expert rating as an explanatory variable. It seems questionable to do this if the aim is to predict the price of wines before they are marketed and, at least in this study, price and rating are not independent variables.

In this study, wines were predominantly from three vintages (2010-2012), and it is possible that more variation would be explained by model fitting if a larger number of vintages were used.

**Acknowledgements**

This study was performed outside my normal working hours. The content is solely the responsibility of the author and does not necessarily represent the official view of either my employer or funding bodies that support my 'normal' research endeavour in genetics. I thank Oliver Mayo and Naomi Wray for helpful comments and Dionysus for inspiration.

Table 1: mean and SD of alcohol percentage, rating and price for different varieties.

|  | Variety | | | | | | | | | |
|---|---|---|---|---|---|---|---|---|---|---|
|  | Red | | | | | White | | | | |
|  | CS | ME | PN | RB | SH | CH | PG | RI | SB | WB |
| Alcohol | 14.0 (0.5) | 13.9 (0.6) | 13.4 (0.5) | 14.0 (0.6) | 14.1 (0.7) | 13.0 (0.5) | 12.8 (0.7) | 11.8 (0.9) | 12.7 (0.7) | 12.4 (0.8) |
| Rating | 93.0 (2.4) | 90.9 (2.5) | 93.1 (2.4) | 92.9 (2.3) | 93.3 (2.3) | 93.2 (2.3) | 91.2 (2.4) | 93.5 (1.9) | 92.0 (2.3) | 91.7 (2.4) |
| Price | 40.6 (38.0) | 27.2 (20.6) | 40.9 (32.2) | 40.1 (58.3) | 41.8 (33.1) | 35.1 (18.7) | 21.6 (7.2) | 24.7 (7.1) | 20.7 (6.2) | 21.5 (9.7) |

Table 2: Results of model fitting

| Variety | Variance explained (%) | |
|---|---|---|
|  | log(Rating) | log(Price) |
| All wines | 12 | 20 |
| Red wines | 7 | 11 |
| White wines | 18 | 27 |
| SH | 4 | 11 |
| RB | 4 | 10 |
| CH | 13 | 10 |
| RI | 9 | 4 |

Table 3: Summary of model fit analyses

| Trait | N | Explanatory variables | $R^2$ model (%) | Reference |
|---|---|---|---|---|
| log(Price) | 2710 | Variety, State, Alcohol, Year | 20 | This study |
| log(Price) | 2710 | + log(Rating) | 53 | This study |
| log(Rating) | 2710 | Variety, State, Alcohol, Year | 12 | This study |
| log(Price) | 519 | Sensory and label characteristics | 63 | [4] |
| log(Rating) | 519 | Sensory and label characteristics | 60 | [4] |
| log(Price) | N/A | Age, weather, technology | 59-78 | [8] |
| log(Price) | N/A | Age, weather | 83 | [3] |
| log(Price) | 6866 | Age, region, rating | 29-45 | [10] |
| log(Price) | 1604 | Region, vintage, producer size, year of marketing, cellaring potential, rating | 58 | [7] |
| $1/\sqrt{}$ (Price) | 559 | Region, variety, producer size, rating | 75 | [5] |

Figure 1: Distribution of the wines by state

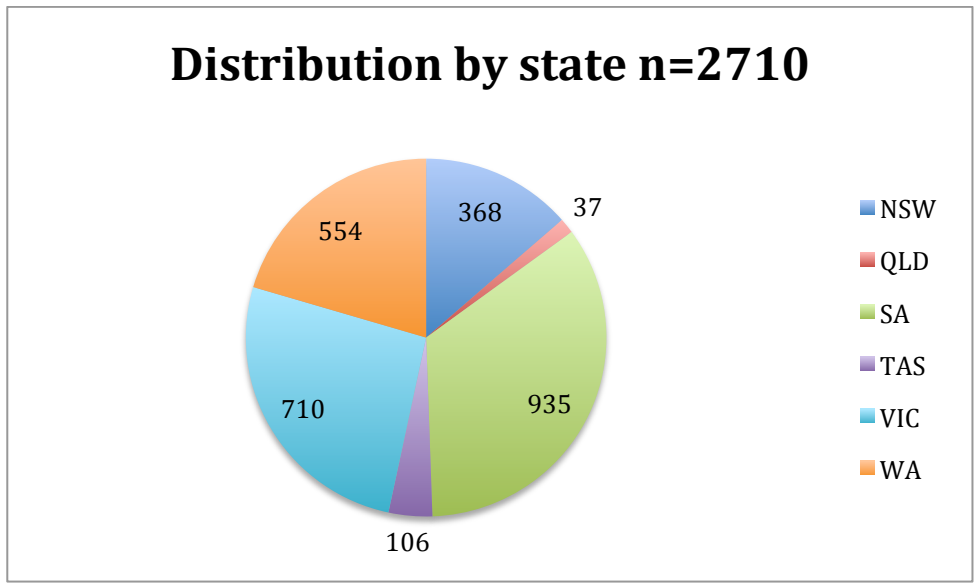

Figure 2: Distribution of the wines by variety

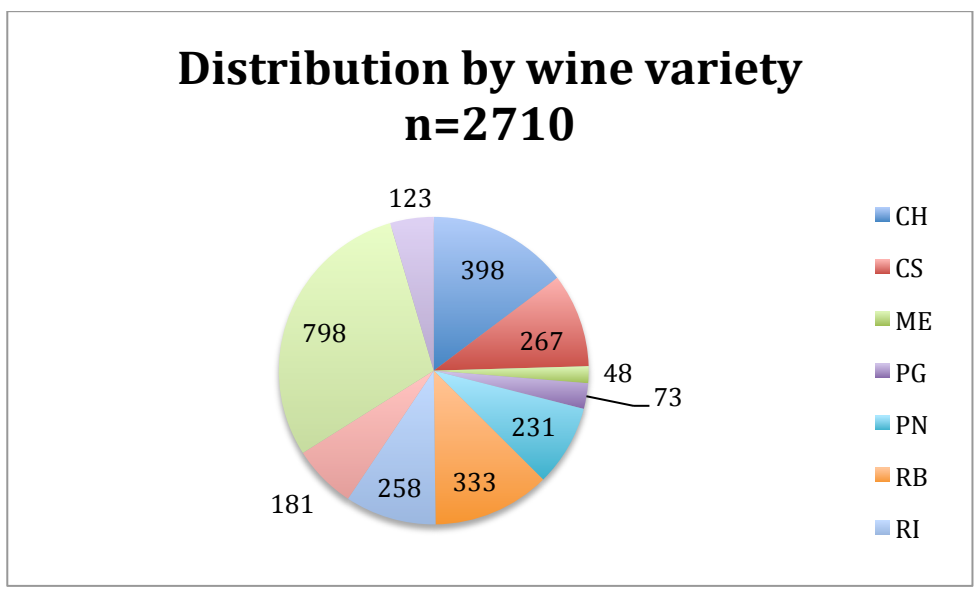

Figure 3: Distribution of wines by State and variety

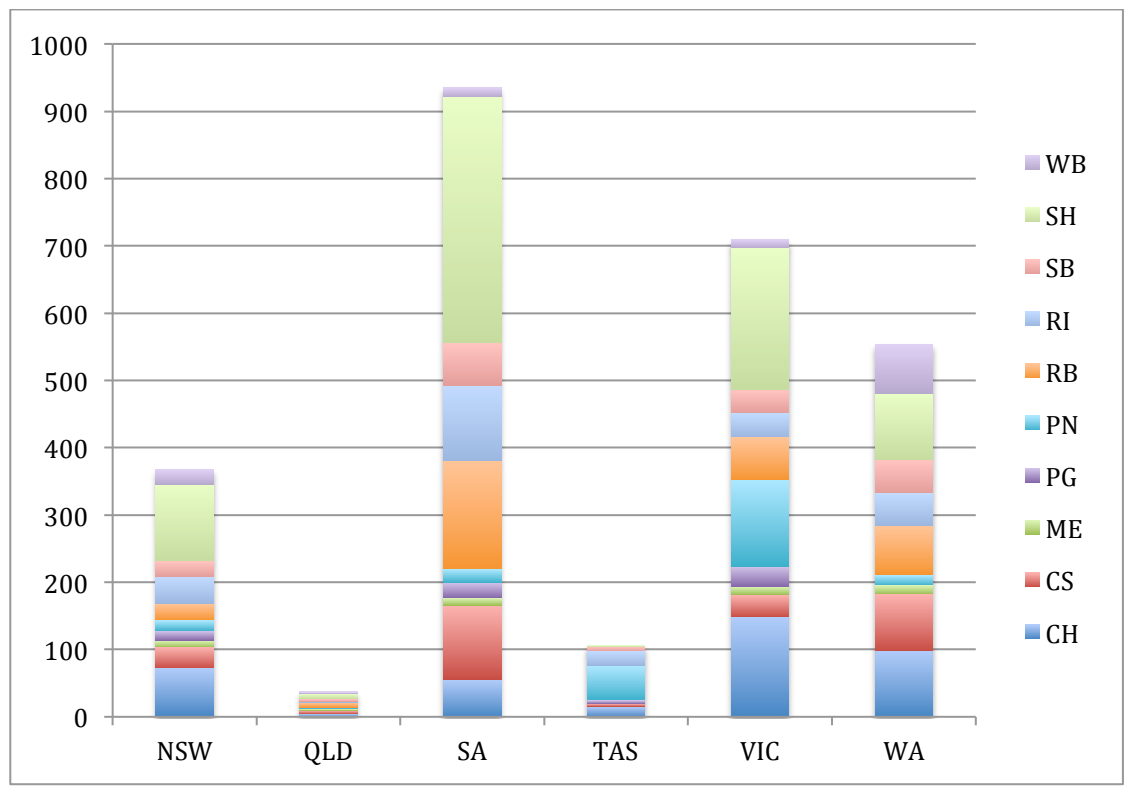

Figure 4: Histogram of year of vintage

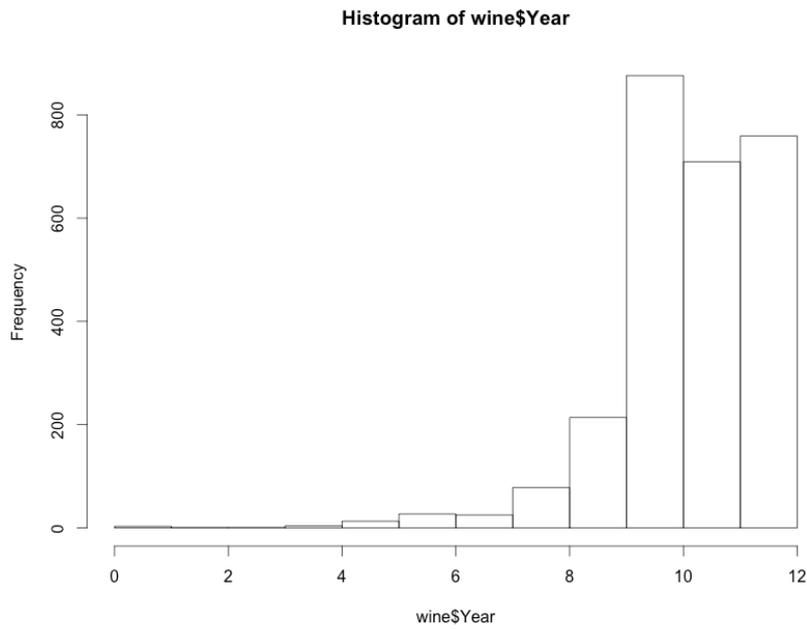

Figure 5: Distribution of alcohol percentage

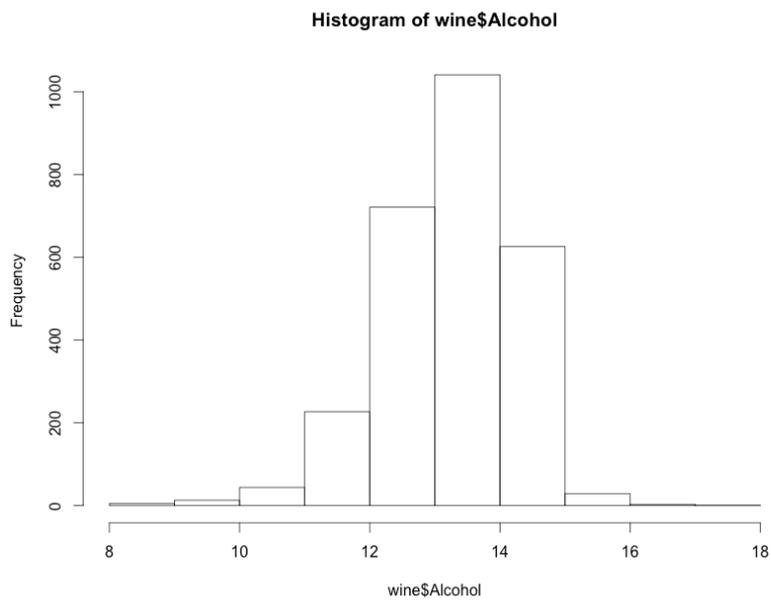

Figure 6: Distribution of log2(Price) by Variety

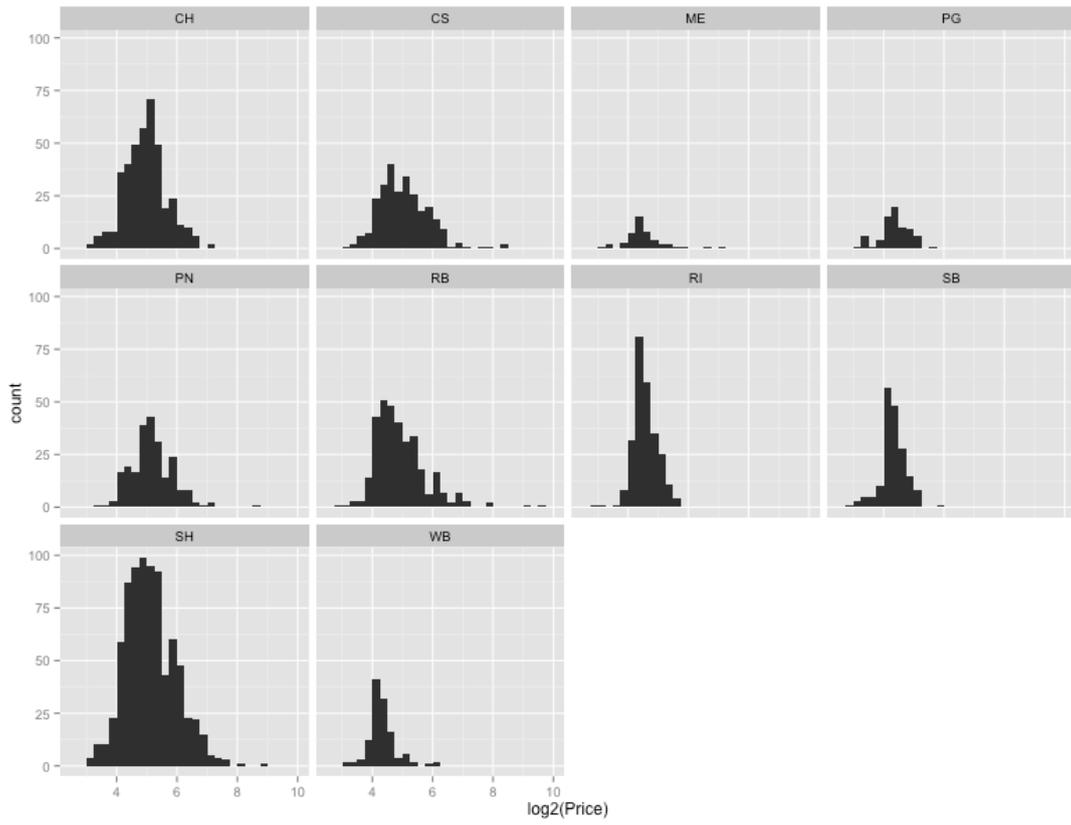

Figure 7: Distribution of Rating by Variety

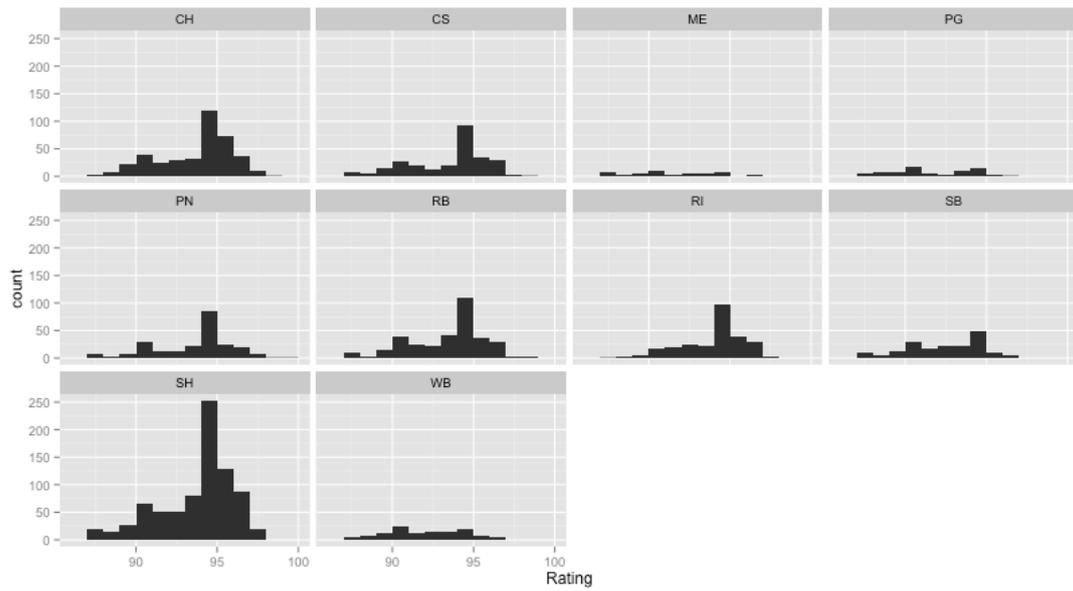

Figure 8: Distribution of log2(Rating) by Variety

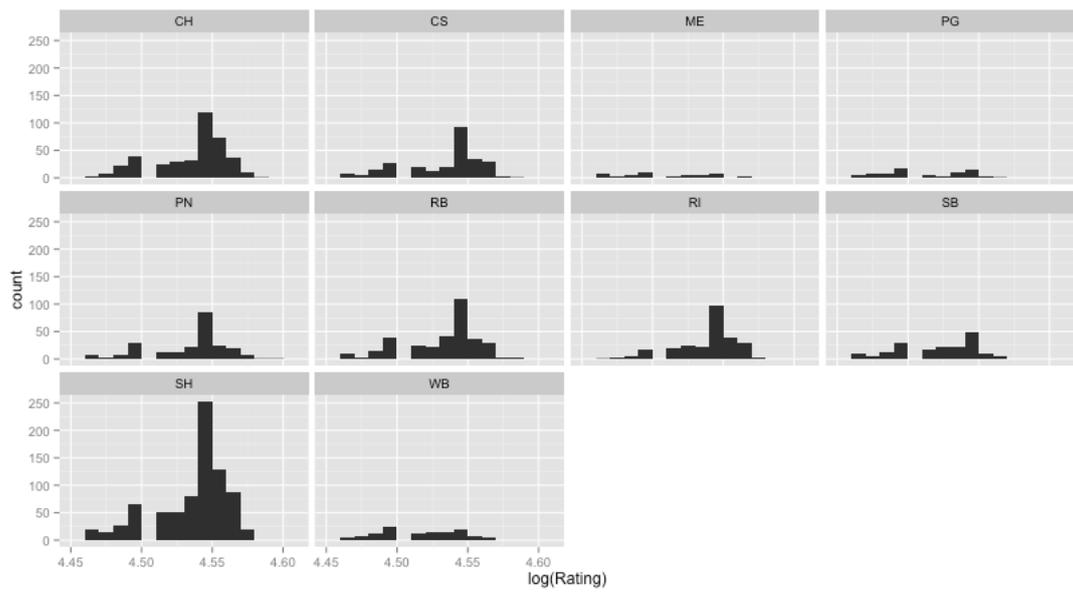

Figure 9: Residuals from model fitting of log2(Rating) against those from a model of log2(Price)

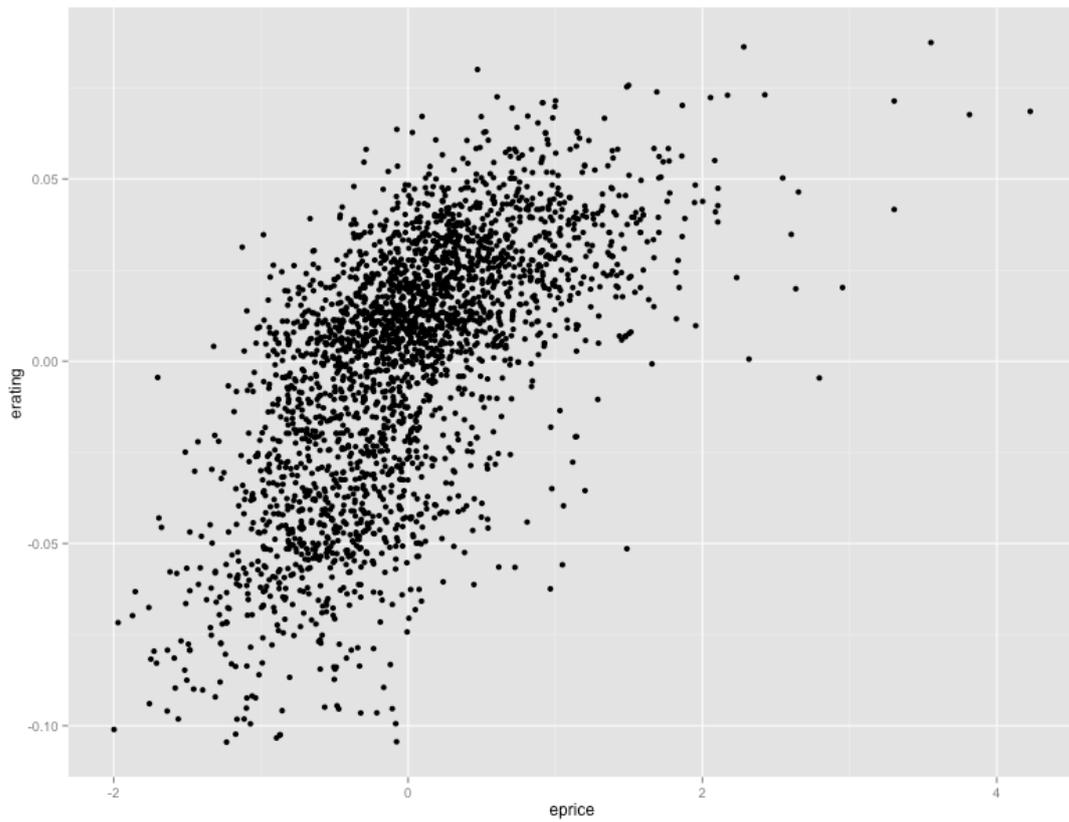

Supplementary Figure 1: Untransformed price against rating

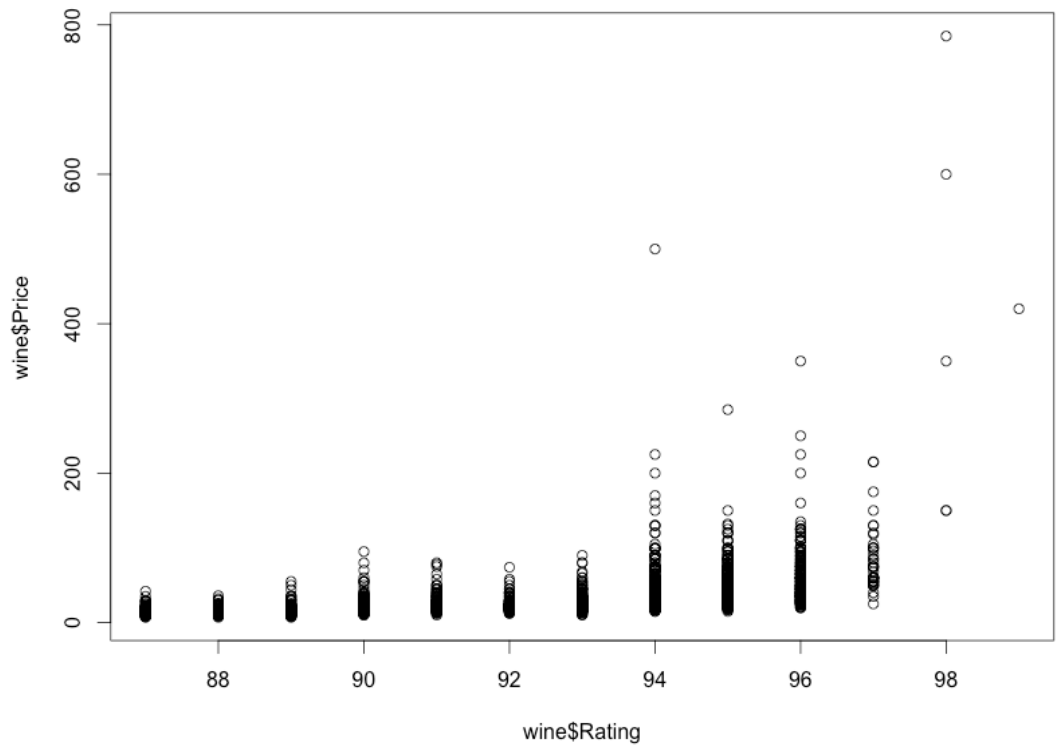